\pdfoutput=1
\documentclass[11pt]{article}
\usepackage{amsmath}
\usepackage{amssymb}
\usepackage{slashed}
\usepackage{graphicx,epstopdf}
\usepackage{cite}
\usepackage{pbox}
\usepackage{geometry}
\usepackage{float}
\usepackage{array}
\usepackage{amsfonts}
\usepackage{graphicx}
\usepackage[font={small}]{caption}
\usepackage[table]{xcolor}
\usepackage{xcolor}
\usepackage{hyperref}
\hypersetup{
    colorlinks=true,
    linkcolor=red,
    filecolor=magenta,      
    urlcolor=blue,
   citecolor=blue,
} 

\providecommand{\xlink}[1]
  {\href{http://arxiv.org/abs/#1}{arXiv:#1}}
\definecolor{RawSienna}{cmyk}{0,0.72,1,0.45}
\definecolor{dgreen}{rgb}{0.0,0.42,0.13}
\definecolor{darkblue}{rgb}{0.0, 0.0, 0.55}
\definecolor{cornellred}{rgb}{0.7, 0.11, 0.11}
\definecolor{calpolypomonagreen}{rgb}{0.08, 0.5, 0.5}

\newcommand{\G}{\color{dgreen}}
\newcommand{\B}{\color{darkblue}}

\def\beq{\begin{equation}}
\def\eeq{\end{equation}}
\def\bea{\begin{eqnarray}}
\def\eea{\end{eqnarray}}
\makeatletter
\@addtoreset{equation}{section}

\begin{document}
\title{\LARGE \bf Consequences of minimal seesaw with complex $\mu\tau$ antisymmetry of neutrinos}
\author{{\bf Rome Samanta$^1$\footnote{ rome.samanta@saha.ac.in}, Probir Roy$^2$\footnote{probirrana@gmail.com}, Ambar Ghosal$^1$\footnote{ambar.ghosal@saha.ac.in},}\\
1. Astroparticle Physics and Cosmology Division\\Saha Institute of Nuclear Physics, HBNI,
  Kolkata 700064, India\\2. Center for Astroparticle Physics and Space Science \\ Bose Institute, Kolkata 700091, India} 
\maketitle
\begin{abstract}
We propose a complex extension of $\mu\tau$ permutation antisymmetry in the neutrino Majorana matrix $M_\nu$. The latter can be realized for the Lagrangian by appropriate CP transformations on the neutrino fields. The resultant form of $M_\nu$ is shown to be simply related to that with a complex (CP) extension of $\mu\tau$ permutation symmetry, with identical phenomenological consequences, though their group theoretic origins are quite different. We investigate those consequences in detail for the minimal seesaw induced by two strongly hierarchical 
right-chiral neutrinos $N_1$ and $N_2$ with the result that the 
Dirac phase is maximal while the two Majorana phases are either 0 or $\pi$. We further provide an uptodate discussion of the $\beta\beta0\nu$ process vis-a-vis ongoing and forthcoming experiments. Finally, a thorough treatment is given of baryogenesis via leptogenesis in this scenario, primarily with the assumption that the lepton asymmetry produced by the decays of $N_1$ only matters here with the asymmetry produced by $N_2$ being washed out. Tight upper and lower bounds on the mass of $N_1$ are obtained from the constraint of obtaining the correct observed range of the baryon asymmetry parameter and the role played by $N_2$ is elucidated thereafter. The mildly hierarchical right-chiral neutrino case (including the quasidegenerate possibility) is discussed in an Appendix.
\end{abstract}
\section{Introduction}\label{s1}
The masses and mixing properties\cite{King:2015aea} of the three light neutrinos continue to intrigue. We now know within reasonably precise ranges their two squared mass differences while from cosmology a fairly tight upper bound\cite{Ade:2015xua} of 0.23 eV has emerged on the sum of the three masses. The atmospheric mixing angle is now pinned around its maximal value of $\pi/4$ and the solar mixing angle around the tri-bimaximal value while the reactor mixing angle is known to be significantly nonzero and close to $8^0$. The current trend of the data\cite{Abe:2017uxa} suggests that the Dirac CP phase could be close to $3\pi/2$ but a definitive statement is yet to emerge. A specific prediction on the value of the latter will be very welcome. It is not known yet whether the light neutrinos are Dirac or Majorana in nature while relentless searches for the decisive neutrinoless double $\beta-$decay signal continue. For the latter case the two Majorana phases of the neutrinos also need to be predicted. Light Majorana neutrino masses can be generated by the seesaw mechanism\cite{seesaw} and a minimal version\cite{King:1999mb} with just two heavy right-chiral (RH) neutrinos seems especially attractive. Further, the formulation of a viable scheme of baryogenesis via leptogenesis within this scenario is a challenging task. There has also been a substantial amount of work with discrete flavor symmetries of the light neutrino Majorana mass matrix: specifically real $\mu\tau$ permutation symmetry\cite{mutaus} and its complex (CP) extension\cite{mutau} as well as real $\mu\tau$ permutation antisymmetry\cite{Grimus:2005jk} but not the complex (CP) extension of that. This last mentioned topic will be the subject of our attention in this paper with the aim of predicting the neutrino CP phases.\\

 The neutrino mass terms in the Lagrangian density read 
\bea
-\mathcal{L}_{mass}^\nu= \frac{1}{2}\bar{\nu_{Ll}^C} (M_\nu)_{lm}\nu_{Lm} + h.c. \label{lag}
\eea 
with $\nu_{Ll}^C=C\bar{\nu_{Ll}}^T$ and the subscripts $l,m$ spanning the lepton flavor indices $e$, $\mu$, $\tau$ while the subscript $L$ denotes left-chiral neutrino fields. $M_\nu$ is a complex symmetric matrix ($M_\nu^*\neq M_{\nu}=M_\nu^T$) in lepton flavor space. It can be put into a diagonal form by a similarity transformation with a unitary matrix $U$:
\bea
U^T M_\nu U=M_\nu^d \equiv \rm diag\hspace{1mm}(m_1,m_2,m_3).\label{e0}
\eea  
Here ${\rm m}_i\hspace{1mm}(i=1,2,3)$ are real and we assume that $m_i\geq 0$. 
We work in the basis in which charged leptons are mass diagonal. We are motivated by a flavor-based model constructed 
by Mohapatra and Nishi\cite{Mohapatra:2015gwa}, which could accommodate a diagonal charged lepton mass matrix as well as a CP-transformed  
$\mu\tau$ interchange symmetry. Now we can relate $U$ to the $PMNS$ mixing matrix $U_{PMNS}$:  
\bea
U=P_\phi U_{PMNS}\equiv 
P_\phi \begin{pmatrix}
c_{1 2}c_{1 3} & e^{i\frac{\alpha}{2}} s_{1 2}c_{1 3} & s_{1 3}e^{-i(\delta - \frac{\beta}{2})}\\
-s_{1 2}c_{2 3}-c_{1 2}s_{2 3}s_{1 3} e^{i\delta }& e^{i\frac{\alpha}{2}} (c_{1 2}c_{2 3}-s_{1 2}s_{1 3} s_{2 3} e^{i\delta}) & c_{1 3}s_{2 3}e^{i\frac{\beta}{2}} \\
s_{1 2}s_{2 3}-c_{1 2}s_{1 3}c_{2 3}e^{i\delta} & e^{i\frac{\alpha}{2}} (-c_{1 2}s_{2 3}-s_{1 2}s_{1 3}c_{2 3}e^{i\delta}) & c_{1 3}c_{2 3}e^{i\frac{\beta}{2}}
\end{pmatrix},\label{eu}
\eea
where $P_\phi={\rm diag}~(e^{i\phi_1},~e^{i\phi_2}~e^{i\phi_3})$ is an unphysical diagonal  phase matrix and  $c_{ij}\equiv\cos\theta_{ij}$, $s_{ij}\equiv\sin\theta_{ij}$ with the mixing angles $\theta_{ij}=[0,\pi/2]$. We work within the PDG convention\cite{Agashe:2014kda} but denote our Majorana phases by $\alpha$ and $\beta$. CP-violation enters through nontrivial values of the Dirac phase $\delta$ and of the Majorana phases $\alpha,\beta$  with $\delta,\alpha,\beta=[0,2\pi]$. \\

Real $\mu\tau$ symmetry \cite{mutaus} for $M_\nu$ implies that 
\bea
G^T M_\nu G=M_\nu,\label{r1}
\eea
where $G$ is a generator of a $\mathbb{Z}_2$ symmetry effecting $\mu\tau$ interchange. In the neutrino flavor space $G$ has the form 
\bea
G=\begin{pmatrix}
1&0&0\\0&0&1\\0&1&0
\end{pmatrix}.\label{gmutau}
\eea 
A substantial amount of phenomenological work has been done following the consequences of (\ref{r1}). Additionally, its possible group theoretic origin from a more fundamental symmetry such as $A_4$ have been investigated\cite{Ishimori:2010au}. However, this flavor symmetry leads to the prediction that $\theta_{13}=0$ which has now been excluded at more than 5.2$\sigma$\cite{An:2015rpe}. A way out was proposed \cite{mutau} in terms of its complex (CP) extension (${\rm CP}^{\mu\tau}$) with the postulate
\bea
G^T M_\nu G=M_\nu^*.\label{mucp}
\eea
The above can be realized as a Lagrangian symmetry by means of a CP transformation on the neutrino fields as
\bea
\nu_{Ll}\rightarrow i G_{lm}\gamma^0\nu_{Lm}^C,
\eea
where $l,m$ are  flavor indices and $\nu_{Lm}^C=C\bar{\nu}_{Lm}^T$. Detailed phenomenological consequences of (\ref{mucp}) have been investigated in Ref.\cite{Mohapatra:2015gwa}.\\

Let us move on to real $\mu\tau$ permutation antisymmetry\cite{Grimus:2005jk} which proposes that
\bea
G^T M_\nu G=-M_\nu.\label{antimu}
\eea
 Note that the antisymmetry condition in (\ref{antimu}) can written as a symmetry condition
\bea
\mathcal{G}^T M_\nu \mathcal{G}=M_\nu,\label{r2}
\eea
where $\mathcal{G}=iG$. Now $\mathcal{G}$ is  a generator of $\mathbb{Z}_4$ symmetry since $\mathcal{G}^4=1$. A sizable amount of work has earlier been done using \cite{Joshipura:2015zla} the real $\mu\tau$ antisymmetry idea -- including its application to the neutrino masses and mixing as well as its possible group theoretic origin from a more fundamental flavor symmetry such as $A_5$. However, the major phenomenological problem with  exact real $\mu\tau$ antisymmetry is that it leads to a maximal solar neutrino mixing angle $\theta_{12}=\pi/4$ as well as two degenerate light neutrinos -- in conflict with experiment \cite{Capozzi:2017ipn}. Perturbative modifications, in attempts to address these problems, unfortunately lead to a proliferation of extra unknown parameters. It is therefore highly desirable to propose an extension of this symmetry which is exact and therefore has the beauty of minimizing the number of input parameters. This is what we aim to do in this paper by proposing a complex (CP) extension of $\mu\tau$ flavor antisymmetry ${\rm CP}^{\mu\tau A}$ and working out its various phenomenological implications.  Complex extensions of $\mu\tau$ symmetry\cite{mutau} and scaling symmetry\cite{CP1} as well as their consequences have been worked out earlier. That is the direction of our thrust here for $\mu\tau$ antisymmetry.\\

 We consider a complex (CP) extension of $\mu\tau$ 
antisymmetry (${\rm CP}^{\mu\tau A}$) in the neutrino mass matrix. 
We show that this extension leads to a form of $M_\nu$ which is very 
simply related to that of $M_\nu$ for the ${\rm CP}^{\mu\tau}$ case. 
Moreover, this form allows neutrino mixing angles that are perfectly 
compatible with experiment both for a normal and for an inverted mass 
ordering. Additionally, specific statements can be made on CP violation 
in the neutrino sector. The Majorana phases $\alpha$ and $\beta$ have to 
be 0 or $\pi$ while Dirac CP violation has to be maximal with the 
phase $\delta$ being either $\pi/2$ or $3\pi/2$. 
Further, reasonably nondegenerate values for the three neutrino masses 
can be generated by incorporating the minimal seesaw 
mechanism \cite{King:1999mb} implemented through two heavy 
right-chiral neutrinos $N_{R \ell}~(\ell=1,2)$ with a $2\times 2$ 
Majorana matrix $M_R$. (In case there is a third heavy Majorana neutrino, 
that is assumed to be much heavier and hence totally decoupled). 
Definitive predictions can be made on neutrinoless double beta decay 
for both types of mass ordering. Finally, a realistic scenario of 
baryogenesis via leptogenesis can be drawn and an acceptable value of the 
baryon asymmetry parameter $Y_B$ can be derived. 
Though the phenomenological consequences of $M_\nu^{{CP}^{\mu\tau}}$ and 
$M_\nu^{{CP}^{\mu\tau A}}$ are identical, we feel that an uptodate 
detailed discussion of these along with some new results related to the scenario of baryogenesis via leptogenesis will be useful.\\

The new features in our work are (i) the demonstration that 
$M_\nu^{{CP}^{\mu\tau A}}$ and $M_\nu^{{CP}^{\mu\tau}}$ have identical phenomenological 
consequences despite there origin from different  residual symmetries 
($Z_4$ and $Z_2$ respectively) and (ii) the use of the minimal seesaw 
with 
two heavy righthanded neutrinos (and consequently one massless 
left handed neutrino) to explore those consequences - in particular 
$\beta\beta0\nu$ decay and baryogenesis via leptogenesis.  
Let us highlight here  what we propose to do in this paper. 
We plan to discuss  the 
complex (CP) extension of $\mu\tau$ antisymmetry  which has been analyzed so far in 
literature  with its perturbative modifications only.  
Then we shall show  how the resultant  
$M_\nu^{{CP}^{\mu\tau A}}$ is simply related to  $M_\nu^{{CP}^{\mu\tau}}$ -- 
the neutrino Majorana mass 
matrix from the complex (CP) extension of $\mu\tau$ symmetry -- with identical 
phenomenological consequences 
despite the fact that their respective real components have almost entirely 
different predictions. 
We further emphasize the fact that ${\rm CP}^{\mu\tau }$ and  
${\rm CP}^{\mu\tau A}$ are 
implemented with different residual symmetry generators, 
namely $\mathbb{Z}_2$ and $\mathbb{Z}_4$ respectively. 
Thus the corresponding high energy theory for these residual 
CP symmetries would likely  be different. 
We then work out the consequences of ${\rm CP}^{\mu\tau A}$ in the 
framework of a minimal seesaw which 
leads to a vanishing value of one of the light neutrino masses and  a very constraint range of the sum of the 
light neutrino masses as well. We also make an uptodate comparison of our 
conclusions on $\beta\beta 0\nu$ decay with ongoing and forthcoming searches. 
We shall do  a full parameter scan of the $3\times 2$ Dirac mass 
matrix $m_D$ in the 
minimal seesaw scenario using the uptodate neutrino oscillation 3$\sigma$ global fit data. 
This in turn will lead us to perform a detailed  computation 
related to the process baryogenesis via 
leptogenesis in our work which will result in new interesting upper and 
lower bounds on the mass of $N_1$. We shall also stress that these bounds could be erased if we consider a mildly hierarchical RH neutrino spectrum.  We shall  discuss the effect of $N_2$ on the final baryon asymmetry $Y_B$, in particular on the obtained upper and lower bounds on $M_1$ from the standard $N_1$ decay scenario.\\

The rest of the paper is organized as follows. In Section \ref{s2} we explain the above mentioned complex extension ${\rm CP}^{\mu\tau A}$.  Section \ref{s3} contains a discussion of how the neutrino mixing angles and CP violating phases originate from  ${\rm CP}^{\mu\tau A}$. In Section \ref{s4} we discuss the origin of the neutrino masses from the minimal seesaw mechanism. The phenomenon of neutrinoless double beta decay is treated in Section \ref{s5}. In Section \ref{s6} we discuss baryogenesis via leptogenesis. Constraints on our model parameter space from all these phenomena are derived by numerical analysis in Section \ref{s7}.   The final Section \ref{s8} contains a discussion of our conclusions. In an Appendix we discuss what 
happens to our results if the right handed neutrinos are mildly hierarchical or quasidegenerate in mass.

\section{Complex extension of $\mu\tau$ antisymmetry}\label{s2}

 We propose a complex extension of (\ref{antimu}), namely
\bea
G^TM_\nu G=-M_\nu^*,~ \mathcal{G}^T M_\nu \mathcal{G}=M_\nu^*.\label{r3}
\eea
The complex invariance condition in (\ref{r3}) can be obtained by the means of a CP transformation\cite{CPt} on the neutrino fields as
\bea
\nu_{Ll}\rightarrow i\mathcal{G}_{lm}\gamma^0\nu_{Lm}^C.
\eea
As we will see, since the real part of the resultant complex matrix exhibits $\mu\tau$ antisymmetry, we call the implemented CP symmetry as a complex extended $\mu\tau$ antisymmetry or simply complex $\mu\tau$ antisymmetry. This complex $\mu\tau$ antisymmetry ${\rm CP}^{\mu\tau A}$, generated by $\mathcal{G}$, needs to be broken in the charged lepton sector. Given that our charged lepton mass matrix $M_\ell$ is diagonal, a replacement of $M_\nu$ by $M_\ell$ in (\ref{r3}) would immediately lead to the unacceptable result $m_\mu=m_\tau$. There is an additional desirable reason for breaking ${\rm CP}^{\mu\tau A}$ in $M_\ell$. A nonzero Dirac CP violation is equivalent to 
\bea
{\rm Tr}~[H_\nu,H_\ell]^3\neq 0, \label{r4}
\eea
where  the hermitian combinations are introduced as $H_\nu=M_\nu^\dagger M_\nu$, $H_\ell=M_\ell^\dagger M_\ell$ \cite{Bernabeu:1986fc}. A common CP symmetry $\mathcal{G}$ in both the sectors would imply
\bea
\mathcal{G}^T H_\nu^T \mathcal{G}^*= H_\nu,~~\mathcal{G}^T H_\ell^T \mathcal{G}^*= H_\ell. \label{r5}
\eea
From (\ref{r5}) it follows that  ${\rm Tr}[H_\nu,H_\ell]^3= 0$  which leads to $\sin \delta =0$ i.e. a vanishing Dirac CP violation. Though this is still a possibility, it goes against the current trend of the data\cite{Abe:2017uxa}. The most general structure of  $M_\nu$ that satisfies the ${\rm CP}^{\mu\tau A}$ condition (\ref{r3}) can be worked out to be 
 \bea
M_\nu^{{CP}^{\mu\tau A}}=\begin{pmatrix}
iA&B&-B^*\\B&C&iD\\-B^*&iD&-C^*
\end{pmatrix},\label{Amutau}
\eea
where $A,D$ are real and $B,C$ are complex mass dimensional quantities which are a priori unknown. The matrix $M_\nu^{{CP}^{\mu\tau A}}$ can also be written as 
\bea
M_\nu^{{CP}^{\mu\tau A}}=\begin{pmatrix}
0&B_1&-B_1\\B_1&C_1&0\\-B_1&0&-C_1
\end{pmatrix}+i\begin{pmatrix}
A&B_2&B_2\\B_2&C_2&D\\B_2&D&C_2
\end{pmatrix},\label{reim}
\eea
where $B=B_1+iB_2$ and $C=C_1+iC_2$  with $B_{1,2}$ and $C_{1,2}$ being real. Note that the real part of the matrix in (\ref{reim}) is invariant under $\mu\tau$ antisymmetry while the imaginary part is $\mu\tau$ symmetric. Thus the entire source of corrections here to  real $\mu\tau$ antisymmetry arises from the imaginary  $\mu\tau$ symmetric part.\\

Here we make the interesting observation that $iM_\nu^{{CP}^{\mu\tau A}}$ yields a neutrino Majorana mass matrix that is complex $\mu\tau$ symmetric since 
\bea
iM_\nu^{{CP}^{\mu\tau A}}=\begin{pmatrix}
-A&iB&-iB^*\\iB&iC&-D\\-iB^*&-D&-iC^*
\end{pmatrix}\equiv M_\nu^{{CP}^{\mu\tau}.}
\eea
This is since
\bea
G^T(i M_\nu^{{CP}^{\mu\tau A}})G=(i M_\nu^{{CP}^{\mu\tau A}})^*.
\eea
Therefore the phenomenological consequences of a complex (CP) $\mu\tau$ symmetric form of $M_\nu$ and a complex antisymmetric form of the same would be identical. Nevertheless, we deem it worthwhile to give a detailed updated discussion of its phenomenological consequences and highlight some new effects such as the role of another heavy RH neutrino $N_2$ on the process of baryogenesis via leptogenesis in a standard $N_1$-leptogenesis scenario.
\section{Neutrino mixing angles and phases from $M_\nu^{{CP}^{\mu\tau A}}$}\label{s3}
 Eqs. (\ref{e0}) and (\ref{r3}) together imply\cite{mutau} that
\bea
\mathcal{G}U^*=U\tilde{d},\label{theorm}
\eea
where 
\bea
\tilde{d}_{lm}=\pm\delta_{lm}.\label{dtld} \eea
Let us take 
\bea
\tilde{d}= \rm diag\hspace{1mm} (\tilde{d_1},\tilde{d_2},\tilde{d_3}),\label{dtl}\eea
where each $\tilde{d}_i$ $(i=1,2,3)$ can be $+1$ or $-1$. Eq. (\ref{theorm}) can   explicitly be written, by taking $\mathcal{G}$ equal to $i$ times $G$ as given in (\ref{gmutau}), namely
\bea
\begin{pmatrix}
i U_{e 1}^*&iU_{e 2}^*&iU_{e 3}^*\\
iU_{\tau 1}^*&iU_{\tau 2}^*&iU_{\tau 3}^*\\
iU_{\mu 1}^*&iU_{\mu 2}^*&iU_{\mu 3}^*
\end{pmatrix}
=\begin{pmatrix}
\tilde{d_1} U_{e 1}&\tilde{d_2}U_{e 2}&\tilde{d_3}U_{e 3},\\
\tilde{d_1}U_{\mu 1}&\tilde{d_2}U_{\mu 2}&\tilde{d_3}U_{\mu 3}\\
\tilde{d_1}U_{\tau 1}&\tilde{d_2}U_{\tau 2}&\tilde{d_3}U_{\tau 3}
\end{pmatrix} \label{Key}
\eea
which is equivalent to six independent equations:
\bea
i U_{e 1}^*=\tilde{d_1} U_{e 1},iU_{e 2}^*=\tilde{d_2}U_{e 2},iU_{e 3}^*=\tilde{d_3}U_{e 3}.\label{1st3}\\
iU_{\tau 1}^*=\tilde{d_1}U_{\mu 1},iU_{\tau 2}^*=\tilde{d_2}U_{\mu 2},iU_{\tau 3}^*=\tilde{d_3}U_{\mu 3}.\label{2nd3}
\eea
\paragraph{}
In order to  calculate the Majorana phases in a way that avoids the unphysical phases, it is useful to construct two rephasing invariants\cite{Branco}
\bea
\mathcal{I}_1=U_{e1}U_{e2}^*,~\mathcal{I}_2=U_{e1}U_{e3}^*.\label{reinv}
\eea
By using (\ref{1st3}), $\mathcal{I}_{1,2}$ can be written  as free of the unphysical phases, namely
\bea
\mathcal{I}_1=\tilde{d}_1\tilde{d}_2U_{e1}^*U_{e2},~\mathcal{I}_2=\tilde{d}_1\tilde{d}_3U_{e1}^*U_{e3}.\label{reinv1}
\eea
After equating the two different expressions for $\mathcal{I}_{1,2}$ in (\ref{reinv}) and (\ref{reinv1}), we obtain
\bea
\mathcal{I}_1=c_{12}s_{12}c_{13}^2e^{-i\frac{\alpha}{2}}=\tilde{d}_1\tilde{d}_2c_{12}s_{12}c_{13}^2e^{i\frac{\alpha}{2}},\label{invp1}\\
\mathcal{I}_2=c_{12}s_{13}c_{13}e^{i(\delta-\frac{\beta}{2})}=\tilde{d}_1\tilde{d}_3c_{12}s_{13}c_{13}e^{-i(\delta-\frac{\beta}{2})}.\label{invp2}
\eea
Eqs.(\ref{invp1}) and (\ref{invp2}) imply 
\bea
e^{i\alpha}=\tilde{d}_1\tilde{d}_2,~e^{2i(\delta-\beta/2)}=\tilde{d}_1\tilde{d}_3
\eea
Thus 
\bea
\tilde{d}_1\tilde{d}_2=+1\Rightarrow\alpha=0,~\tilde{d}_1\tilde{d}_2
=-1\Rightarrow\alpha=\pi, \label{alph1}\\
\tilde{d}_1\tilde{d}_3=+1\Rightarrow\delta-\frac{\beta}{2}=0,~\tilde{d}_1\tilde{d}_3
=-1\Rightarrow\delta-\frac{\beta}{2}=\pi/2.\label{beta1}
\eea
\paragraph{}
Taking the modulus squared of the third equality in (\ref{2nd3}), namely $|U_{\tau 3}|=|U_{\mu 3}|$, we obtain
\bea
c_{23}^2=s_{23}^2\label{maxth23}
\eea
which implies $\theta_{23}=\pi/4$, i.e. a maximal atmospheric mixing. Incorporating this last result, the modulus square of the first or the second equality in (\ref{2nd3}) leads after some algebra to the relation
\bea
2c_{12}s_{12}c_{13}s_{13}\cos\delta=0.\label{cd0}
\eea
Given the experimentally observed nonvanishing values for all the mixing angles, (\ref{cd0}) leads to a maximal Dirac CP-violation
\bea
\cos\delta=0~i.e.~\delta=\pi/2~{\rm or} ~3\pi/2.\label{cosdl}
\eea
It then follows from (\ref{beta1}) that
\bea
\tilde{d}_1\tilde{d}_3=+1\Rightarrow\beta=\pi,~\tilde{d}_1\tilde{d}_3
=-1\Rightarrow\beta=0.\label{beta}
\eea
We can summarize our results on $\alpha$, $\beta$ and $\cos\delta$ in Table \ref{t1}.
\begin{table}[H]
\begin{center}
\caption{Predictions on the CP phases} \label{t1}
 \begin{tabular}{|c|c|c|c|c|c|c|c|} 
\hline 
$\tilde{d}_1$&$\tilde{d}_2$&$\tilde{d}_3$&$\tilde{d}_1\tilde{d}_2$&$\tilde{d}_1\tilde{d}_3$&$\alpha$ &$\beta$ &$ \cos \delta$ \\
\hline
1&1&1&1&1&$0$&$\pi$&$0$\\
1&-1&1&-1&1&$\pi$&$\pi$&$0$\\
1&1&-1&1&-1&$0$&$0$&$0$\\
1&-1&-1&-1&-1&$\pi$&$0$&$0$\\
\hline 
\end{tabular} 
\end{center} 
\end{table}
\section{Origin of neutrino masses from a minimal  seesaw}\label{s4}
\noindent
We now discuss  the realization of the complex extended $\mu\tau$ 
antisymmetric  mass matrix $M_\nu^{{CP}^{\mu\tau A}}$ through the minimal  
seesaw mechanism\cite{King:1999mb} mentioned earlier. This mechanism makes 
use of two heavy right chiral neutrino fields $N_{Ri}\hspace{.5mm}(i=1,2)$ 
with a Majorana mass matrix $M_R$. We work in a basis in which $M_R$ is real, 
positive and diagonal\cite{Chen:2016ptr}, i.e.,  
$M_R=\rm diag\hspace{.5mm} (M_1,M_2)$, $M_{1,2}>0$. 
With $m_D$ as the Dirac mass matrix, the neutrino mass terms read 
\bea
-\mathcal{L}_{mass}^{\nu,N}= \bar{N}_{Ri} (m_D)_{i\alpha}l_{L\alpha}
+\frac{1}{2}\bar{N}_{Ri}(M_R)_{ij} \delta _{ij}N_{Rj}^C 
+ {\rm h.c}., \label{seesawlag}
\eea
where $l_{L\alpha}=\begin{pmatrix}\nu_{L\alpha} & e_{L\alpha}\end{pmatrix}^T$ is the SM lepton doublet of flavor $\alpha$. The effective light neutrino mass matrix is given by the standard seesaw relation
\bea
M_\nu = -m_D^TM_R^{-1}m_D. \label{seesaweq}
\eea
In this case (\ref{r3}) is satisfied through the symmetry transformation on $m_D$ as 
\bea
m_D\mathcal{G}=-im_D^*,\label{mdtr}
\eea
so long as $M_R^{-1}$ is real. The most general form of $m_D$ that satisfies (\ref{mdtr}) can be parametrized as
\bea
m_D=\begin{pmatrix}
\sqrt{2}a_1e^{i\pi/4}&b_1e^{i\theta_1}&ib_1e^{-i\theta_1}\\
\sqrt{2}a_2e^{i\pi/4}&b_2e^{i\theta_2}&ib_2e^{-i\theta_2}\\
\end{pmatrix}, \label{mdseesaw}
\eea
where the  parameters $a_{1,2}$, $b_{1,2}$ and $\theta_{1,2}$  are real.\\

The form of the effective light neutrino mass matrix $M_\nu$ that now emerges is given below:
\bea
M_\nu^{{CP}^{\mu\tau A}}=\hspace{16cm}\nonumber\\
\begin{pmatrix}
-2i(x_1^2+x_2^2)&-\sqrt{2}e^{i\pi/4}(x_1y_1e^{i\theta_1}+x_2y_2e^{i\theta_2})&-i\sqrt{2}e^{i\pi/4}(x_1y_1e^{-i\theta_1}+x_2y_2e^{-i\theta_2})\\-\sqrt{2}e^{i\pi/4}(x_1y_1e^{i\theta_1}+x_2y_2e^{i\theta_2})&-(e^{2i\theta_1}y_1^2+e^{2i\theta_2}y_2^2)&-i(y_1^2+y_2^2)\\-i\sqrt{2}e^{i\pi/4}(x_1y_1e^{-i\theta_1}+x_2y_2e^{-i\theta_2})&-i(y_1^2+y_2^2)&e^{-2i\theta_1}y_1^2+e^{-2i\theta_2}y_2^2
\end{pmatrix}.\nonumber\\
\nonumber\eea
\bea
\hspace{16cm}\label{mnuseesaw}
\eea
In (\ref{mnuseesaw}) we have introduced new real parameters $x_{1,2}$ and $y_{1,2}$ which are obtained by scaling $a_{1,2}$ and $b_{1,2}$ with the square roots of the respective $RH$ neutrino masses $M_{1,2}$, i.e.
\bea
\frac{a_{1,2}}{\sqrt{M_{1,2}}}= x_{1,2},~\frac{b_{1,2}}{\sqrt{M_{1,2}}}= y_{1,2}.\label{primed}
\eea
The lightest neutrino mass, either $m_1$ for a normal mass ordering or $m_3$ for an inverted mass ordering, has to vanish since ${\rm det}~M_\nu^{{CP}^{\mu\tau A}}=0$. Furthermore, one of the phases of $M_\nu$ (say $\theta_1$) can be rotated by the phase matrix $P_\phi={\rm diag}~(1,e^{i\phi},e^{-i\phi})$ with the choice $\theta_1=-\phi$. Thus we are left with only the phase difference $\theta_2-\theta_1$ in $M_\nu$. We can now rename $\theta_2-\theta_1$ as $\theta$. Without  loss of generality, this is also equivalent to the choice $\theta_1=0$ and  $\theta_2=\theta$ in $m_D$. From now on we shall use this redefined phase $\theta$ for both  $M_\nu$ and $m_D$. 
\noindent
\section{Neutrinoless double beta decay}\label{s5}
The rare $\beta\beta0\nu$ process can arise from the following decay of of a nucleus 
\bea
(A,Z)\longrightarrow (A, Z+2)+2e^-.\label{betadec}
\eea 
In (\ref{betadec})  lepton number is violated by two units. Unlike in neutrinoful double $\beta-$decay, which is a sequence of two single $\beta-$decays, final state neutrinos are absent in the $\beta\beta 0\nu$ process. The latter can go through via an appropriate neutrino loop only if the light neutrinos have Majorana masses. Therefore any  observation of such a decay will unambiguously  establish the Majorana nature of the light neutrinos. The half-life, corresponding to $\beta\beta 0\nu$ decay, can be expressed as
\bea
\frac{1}{T^{0\nu}_{1/2}}=G|M_{ee}|^2|\mathcal{M}|^2 m_e^{-2}. 
\eea 
Here $G$ is the two-body phase space factor and $M_{ee}$ is the (1,1) element of the effective light neutrino mass matrix $M_\nu$, cf.(\ref{lag}). Moreover, $\mathcal{M}$ is the nuclear matrix element (NME) and $m_e$ is the electron mass.  $M_{ee}$ can be written within our convention as
\bea
M_{ee}=c_{12}^2c_{13}^2m_1+s_{12}^2c_{13}^2m_2e^{i\alpha}+s_{13}^2m_3e^{i(\beta-2\delta)}.\label{betad}
\eea
\paragraph{}
Significant upper limits on $|M_{ee}|$ are available from  ongoing search experiments for $\beta\beta 0\nu$ decay. KamLAND-Zen \cite{Asakura:2015ajs} and EXO \cite{Auger:2012ar} had earlier constrained this value to be $<0.35$ eV. But the most impressive upper bound till date  is provided by  GERDA phase-II data\cite{Majorovits:2015vka}: $M_{ee}<0.098$ eV. As explained in Sec.\ref{s3}, we have four sets of values for the three CP violating phases $\alpha,\beta,\delta$ in the neutrino sector corresponding to the four independent $\tilde{d}$ matrices. Furthermore, we need to consider both kinds of light neutrino mass ordering: normal and inverted. Thus we shall have eight sets of predictions for $|M_{ee}|$ from our modelled $M_\nu$. These will be detailed in our section on numerical analysis.\\

At this stage it may be useful to point out how (\ref{betad}) simplifies in 
our model for the specific cases of normal and inverted mass ordering subject 
to the condition given in eqn.(3.16). 
For a normal mass ordering, we have $m_1=0$ and further 
\bea
\alpha=0,\beta=0;~\alpha=\pi,\beta=\pi:~|M_{ee}|=\left(s_{12}^4c_{13}^4 m_2^2+s_{13}^4m_3^2-2s_{12}^2s_{13}^2c_{13}^2m_2m_3\right)^{1/2}({\rm Normal}),\\
\alpha=0,\beta=\pi;~\alpha=\pi,\beta=0:~|M_{ee}|=\left(s_{12}^4c_{13}^4 m_2^2+s_{13}^4m_3^2+2s_{12}^2s_{13}^2c_{13}^2m_2m_3\right)^{1/2}({\rm Normal}).
\eea
 Note that the value of $|M_{ee}|$ becomes somewhat less here since the terms involving $m_3$ are suppressed by the powers of $s_{13}$. For an inverted mass ordering, $m_3$=0 and $|M_{ee}|$ becomes independent of $\beta$ and $\delta$. Indeed, we have
 \bea
 \alpha=0:~|M_{ee}|=c_{13}^2\left(c_{12}^4m_1^2+s_{12}^4m_2^2+2c_{12}^2s_{12}^2m_1m_2\right)^{1/2}(\rm Inverted),\\
 \alpha=\pi:~|M_{ee}|=c_{13}^2\left(c_{12}^4m_1^2+s_{12}^4m_2^2-2c_{12}^2s_{12}^2m_1m_2\right)^{1/2}(\rm Inverted).
 \eea
 Since $\Delta m^2_{21}<<|\Delta m^2_{32}|$, in this case we can assume $m_1\approx m_2\approx \sqrt{|\Delta m_{32}^2|}$.
Thus, for the two allowed values of $\alpha$, we have 
\bea
 \alpha &=& 0:~|M_{ee}| \simeq \sqrt{|\Delta m_{32}^2|}c_{13}^2~ (\rm Inverted),\\
 \alpha &=&\pi :~|M_{ee}|\simeq \sqrt{|\Delta m_{32}^2|}c_{13}^2 [\lbrace 1-2s_{12}^2\rbrace ^2 ]~(\rm Inverted).
\eea
We see that $|M_{ee}|$ for $\alpha=\pi$ is suppressed here relative to its value in the $\alpha=0$ case.
\section{Baryogenesis via leptogenesis}\label{s6}
To start with, we recall the observed range of $Y_B=(n_B-n_{\bar{B}})/s$ $-$ the ratio of baryonic minus antibaryonic number density to the entropy density -- namely 
\bea
8.55\times 10^{-11}<Y_B<8.77\times 10^{-11}. \label{Ybob}
\eea
CP violating  decays from heavy Majorana neutrinos that are out of equilibrium generate a lepton asymmetry\cite{Fukugita:1986hr,Riotto:1999yt,Davidson:2008bu}. The latter is later converted into a baryon asymmetry by sphaleron transitions\cite{Kolb:1990vq}. The appropriate part of the  Lagrangian for the process can be written as
\bea
 -\mathcal{L}=\lambda_{i\alpha} \bar{N}_{Ri}\tilde{\phi}^\dag {l}_{L\alpha}+\frac{1}{2}\bar{N}_{Ri}(M_R)_{ij} \delta _{ij}N_{Rj}^C+{\rm h.c.},\label{cpa}
 \eea
 where $\tilde{\phi}=i\tau_2\phi^*$, with $\phi=\begin{pmatrix}\phi^+ & \phi^0\end{pmatrix}^T$ being the Higgs doublet. The possible decays of $N_i$ from \eqref{cpa} are $N_i\to e^-_\alpha\phi^+$, $\nu_\alpha\phi^0$, $ e_\alpha^+\phi^-$, and $ \nu^C_\alpha \phi^{0*}$. 
The CP asymmetry parameter $\varepsilon_i^\alpha$,  that is a measure of the required CP violation, arises from the interference between the tree level, one loop self energy and one loop vertex  diagrams\cite{Fukugita:1986hr} for the decay of $N_i$. It has the general expression\cite{Pilaftsis:2003gt}
\begin{equation}
\varepsilon_i^\alpha=\frac{1}{4\pi v^2 h_{ii}}\sum\limits_{j\neq i}\left\{\text{I}m[h_{ij}(m_D)_{i\alpha}(m_D^*)_{j\alpha}]g(x_{ij})+\frac{\text{Im}[h_{ji}(m_D)_{i\alpha}(m_D^*)_{j\alpha}]}{1-x_{ij}}\right\},\label{asymp}
\end{equation} 
where $h_{ij}\equiv( m_D m_D^\dag)_{ij}$, $\langle\phi^0\rangle=v/\sqrt{2}$ (so that $m_D = v\lambda/\sqrt{2}$) and $x_{ij}=M_j^2/M_i^2$. In addition, the loop function
$g(x_{ij})$ has the standard expression
\begin{equation}
g(x_{ij})=\frac{\sqrt{x_{ij}}}{1-x_{ij}}+f(x_{ij})\label{cpa2}
\end{equation} 
with
\begin{equation}
f(x_{ij})=\sqrt{x_{ij}}\Big[1-(1+x_{ij})\ln\Big(\frac{1+x_{ij}}{x_{ij}}\Big) \Big].
\end{equation} 

Before proceeding further in the calculation of 
$\varepsilon_i^\alpha$ in our scenario, we need to address some 
important issues related to 
leptogenesis. For hierarchical  RH neutrino masses $M_2\gg M_1$ 
(some discussion of the mildly hierarchical RH neutrino case including 
quasidegenerate masses is given later in the Appendix), it can be shown that only the decays of $N_1$ matter for the creation of lepton asymmetry while the latter created from the heavier neutrinos gets washed out\cite{Buchmuller:2003gz} significantly. Therefore, in general, only $\varepsilon_1^\alpha$ is the pertinent quantity in a hierarchical leptogenesis scenario. Nevertheless, there are certain circumstances in which the decays of $N_{2,3}$ do affect the final baryon asymmetry\cite{barin2,Engelhard:2006yg}. Furthermore, flavor plays an important role in the phenomenon of leptogenesis\cite{Abada:2006ea,Abada:2006ea2}. Assuming the temperature scale of the process to be $T\sim M_1$,  the rates of the charged lepton Yukawa interaction categorize leptogenesis  into the following three categories. \\

\noindent
1) {\it Unflavored leptogenesis}: $T\sim M_1>10^{12}$ GeV, when all interactions with all flavors are out of equilibrium: In this case all the flavors are indistinguishable; therefore the total CP asymmetry is a sum over all flavors, i.e. $\varepsilon_1=\sum_{\alpha}\varepsilon^\alpha_1$ and the final baryon asymmetry $Y_B$ is proportional to $\varepsilon_1$.\\ 

\noindent
 2) {\it  $\tau$-flavored leptogenesis:} $10^9$ GeV $<T\sim M_1 <10^{12}$ GeV, when only the $\tau$ flavor is in equilibrium and hence distinguishable. In this regime there are two pertinent CP asymmetry parameters; $\varepsilon_1^\tau$ and $\varepsilon_1^{(2)}=\varepsilon_1^e+\varepsilon_1^\mu$. The final baryon asymmetry $Y_B$ may be approximated as\cite{Abada:2006ea}
\bea
Y_B\simeq-\frac{12}{37g^*}\Big[\varepsilon_1^{(2)}\eta\Big(\frac{417}{589}\tilde{m}_2\Big) + \varepsilon_1^{\tau}\eta\Big(\frac{390}{589}\tilde{m}_\tau\Big)\Big],\label{tflv}
\eea where the washout masses $\tilde{m}_{2,\tau}$ and $\varepsilon_1^{(2)}$ are defined as 
\bea
\tilde{m}_2=M_1^{-1}\left(|(m_D)_{1e}|^2+|(m_D)_{1\mu}|^2\right), \hspace{1mm}\tilde{m}_\tau= M_1^{-1}|(m_D)_{1\tau}|^2,\hspace{1mm}
\varepsilon_1^{(2)}=\sum\limits_{\alpha=e,\mu}\varepsilon_1^{\alpha}=\varepsilon_1^{e}+\varepsilon_1^{\mu}.
\eea 
In order to know the nature of the washout processes, it is convenient to define  two washout parameters $K_{2,\tau}=\tilde{m}_{2,\tau}/10^{-3}$ relevant to this mass regime. Further, $\eta(\tilde{m}_{2})$ and  $\eta(\tilde{m}_{\tau})$ are the efficiency factors that account for the inverse decay and the lepton number violating scattering processes while $g^*$ is the number of relativistic degrees of freedom in the thermal bath having a value $g^*\approx 106.75$ in the SM. \\
 
\noindent 
3) {\it Fully flavored leptogenesis:} $T\sim M_1<10^9$ GeV, when in addition to the $\tau$ flavor, the $\mu$ flavor is also in equilibrium $-$ thus all the three flavors are distinguishable. Again for the evaluation of the final baryon asymmetry $Y_B$ in this regime, we make use of the approximate analytic formula for $Y_B$ presented in Ref.\cite{Abada:2006ea}. In the $T\sim M_1<$ $10^9$ GeV regime, $Y_B$ is well approximated by
\bea
Y_B\simeq-\frac{12}{37g^*}\Big[\varepsilon_1^{e}\eta\Big(\frac{151}{179}\tilde{m}_e\Big) + \varepsilon_1^{\mu}\eta\Big(\frac{344}{537}\tilde{m}_\mu\Big)+\varepsilon_1^{\tau}\eta\Big(\frac{344}{537}\tilde{m}_\tau\Big) \Big],
\label{fflv}\eea 
 where the washout masses $\tilde{m}_\alpha$ are defined as
\bea
\tilde{m}_\alpha=\frac{|(m_D)_{1\alpha}|^2}{M_1},~{\alpha=e,\mu,\tau}.
\eea
We now focus on the calculation of the quantities related to the leptogenesis in our model. The flavor sum over $\alpha$ leads  the first term in the RHS of \eqref{asymp} to be proportional to ${\rm Im} (h_{ij})^2$ and  the  second term to vanish. This is since
\begin{equation}
\sum\limits_{\alpha}\text{Im}[h_{ji}(m_D)_{i\alpha}(m_D^*)_{j\alpha}]=\text{Im}[h_{ji}h_{ij}]=\text{Im}[h_{ji}h_{ji}^*]=\text{Im}|h_{ji}|^2=0.
\end{equation} 
In fact, in our model the matrix $h=m_Dm_D^\dagger$ is real as given by
\bea
h=\begin{pmatrix}
2(a_1^2+b_1^2)&2 (a_1 a_2 + b_1 b_2 \cos \theta)\\2 (a_1 a_2 + b_1 b_2 \cos \theta)&2(a_2^2+b_2^2)
\end{pmatrix}.\label{hmat}
\eea
Therefore the flavor summed CP asymmetry parameter $\varepsilon_1=\sum_{\alpha}\varepsilon^\alpha_1$ vanishes, i.e.,  unflavored leptogenesis does not occur in this complex (CP) extended $\mu\tau$ antisymmetry scheme. Using (\ref{mdseesaw}) and (\ref{asymp}), the flavored CP asymmetries can be calculated to be
\bea
\varepsilon_1^e=0,\hspace{1mm}\varepsilon_1^\mu=-\frac{g^\prime(x_{12})}{4\pi v^2}\left[\frac{(a_1a_2+b_1b_2\cos\theta)b_1b_2\sin\theta}{a_1^2+b_1^2}\right]=-\varepsilon_1^\tau,\label{CPmod}
\eea
where $g^\prime(x_{12})$ is given by
\bea
g^\prime(x_{12})=g(x_{12})+(1-x_{12})^{-1}.\label{gpr}
\eea
It  is  useful to simplify (\ref{gpr}) for a hierarchical RH neutrino scheme to
\bea
g^\prime(M_2^2/M_1^2)=-\frac{3}{2}\frac{M_1}{M_2}-\frac{M_1^2}{M_2^2}.\label{m1m2}
\eea
Now in our minimal seesaw scheme, assuming a specific hierarchy of the RH  neutrino masses, namely $M_2/M_1\simeq 10^3$, the final $Y_B$  is calculated from (\ref{CPmod}), (\ref{tflv}) and (\ref{fflv}) to be
\bea
Y_B\simeq \frac{12}{37g^*} \varepsilon_1^{\mu}\Big[ \eta\Big(\frac{390}{589}\tilde{m}_\tau\Big)-\eta\Big(\frac{417}{589}\tilde{m}_2\Big) \Big]\label{YBtau}
\eea
for the $\tau-$flavored regime and 
\bea
Y_B\simeq \frac{12}{37g^*} \varepsilon_1^{\mu}\Big[ \eta\Big(\frac{344}{537}\tilde{m}_\tau\Big)-\eta\Big(\frac{344}{537}\tilde{m}_\mu\Big) \Big]\label{YBsign}
\eea
for the fully flavored regime. \\

In our primary  analysis, the effect of the  heavy neutrino ($N_2$) on the produced final baryon asymmetry has been neglected with the assumption that the asymmetry produced by the decays of $N_2$ get  washed out\cite{Buchmuller:2003gz}. We now give a brief discussion  on how the heavy neutrino $N_2$ can affect the final baryon asymmetry $Y_B$.  As elaborated below, there are two ways in which the effect of $N_2$  might arise: indirect and direct. We first discuss the {\it \G indirect effect}. Though the neutrino oscillation data are fitted with the rescaled parameters of (\ref{primed}), in order to compute  the quantities related to leptogenesis such as $\varepsilon_1^\alpha$, we need to evaluate the  parameters of the Dirac mass matrix elements. Given a set of rescaled parameters, the latter can be generated by varying $M_{1,2}$ in (\ref{primed}). It is thus interesting to see whether the final baryon asymmetry is affected by the chosen mass ratios of the RH neutrinos. We find that  the final $Y_B$ is not particularly sensitive to $M_{2}$. A  relook at (\ref{m1m2}) reminds us that the second term is suppressed compared to the first term, since the former is of the order of $x_{12}^{-1}$. Thus, taking only the first term of (\ref{m1m2}) into consideration, the flavored CP asymmetry parameters of (\ref{CPmod}) can be simplified in terms of the rescaled parameters of (\ref{primed}) as
\bea
\varepsilon_1^\mu=\frac{3M_1}{8\pi v^2}\left[\frac{(x_1x_2+y_1y_2\cos\theta)y_1y_2\sin\theta}{x_1^2+y_1^2}\right]=-\varepsilon_1^\tau. \label{epssim}
\eea  

 Since all  rescaled parameters in (\ref{epssim}) are fixed by the $3\sigma$ oscillation data, $\varepsilon_1^{\mu,\tau}$ are practically insensitive to the value of $M_2$. Nevertheless, for a precise numerical computation of the final baryon asymmetry, we need to take into account the effect of the second term in (\ref{m1m2}). The sensitivity of $Y_B$ to the magnitude of the second term of (\ref{m1m2}) for different mass hierarchical schemes of the RH neutrinos will be discussed in detail in the numerical Section \ref{s7}.\\

We now turn to discuss the {\it \G direct effect} of $N_2$. We have so far focused on the lepton asymmetry produced by the decay of the lightest of the heavy neutrinos. It is shown in Ref.\cite{Engelhard:2006yg} that, due to a decoherence effect, the amount of lepton asymmetry, generated by $N_2$ decays,  gets protected against  $N_1$-washout. The latter therefore survives down to the electroweak scale and  contributes to the final
 baryon asymmetry. For this procedure to work out, two washout parameters $\Delta_{1}={h_{11}}{M_1^{-1}{m^*}^{-1}}$ and $\Delta_2={h_{22}}{M_2^{-1}{m^*}^{-1}}$ must satisfy the condition 
 \bea
 \Delta_1\gg 1\hspace{1mm} {\rm and}\hspace{1mm}\Delta_2 \not\gg 1\label{n2cond}
 \eea
  with  $m^*=1.66\sqrt{g^*}\pi v^2/M_{Pl}\approx 10^{-3}$ eV. Here $\Delta_1\gg 1$ indicates that very fast $N_1$ interactions destroy the coherence among the states produced by $N_2$; hence a part of the lepton asymmetry produced by $N_2$  becomes blind to the $N_1$-washout and survives orthogonal to $N_1$-states. On the other hand, a mild washout of the lepton asymmetry, produced by $N_2$ due to $N_2$-related interactions, is represented by the  $\Delta_2 \not\gg 1$ condition. For such a mild washout scenario,  a sizable  lepton asymmetry generated by $N_2$ survives through the $N_1$-leptogenesis phase and hence contributes to the final baryon asymmetry. We shall elaborate on the validity of these conditions in our model in the following section.
\section{Numerical analysis: methodology and discussion} \label{s7} 
\begin{figure}[H]
\begin{center}

\includegraphics[scale=.45]{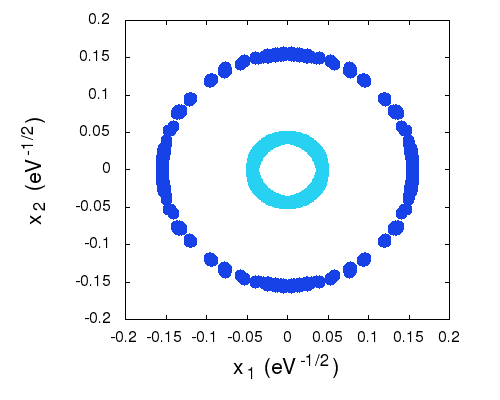}\includegraphics[scale=.45]{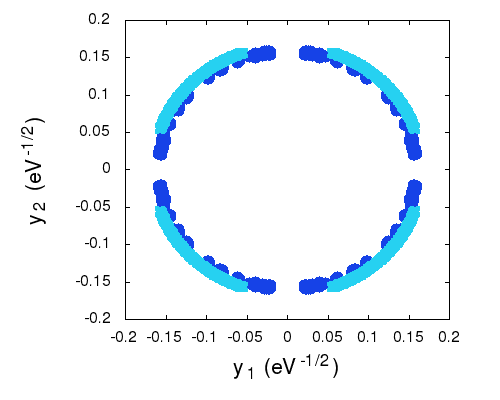}\\\includegraphics[scale=.45]{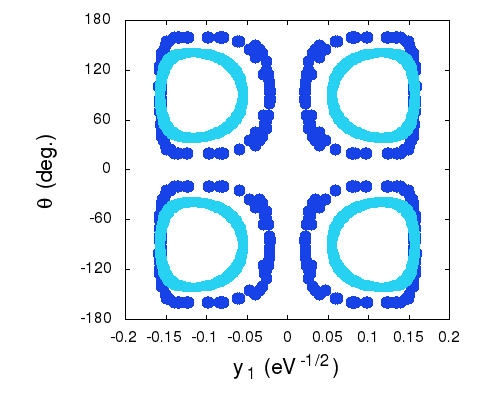}
\end{center}
\caption{\it Rescaled parameter space for both the mass orderings. The plots in sky blue (deep blue) color represents the parameter space for normal (inverted) mass ordering. }\label{fig1}
\end{figure}

\begin{figure}[H]
\includegraphics[scale=.22]{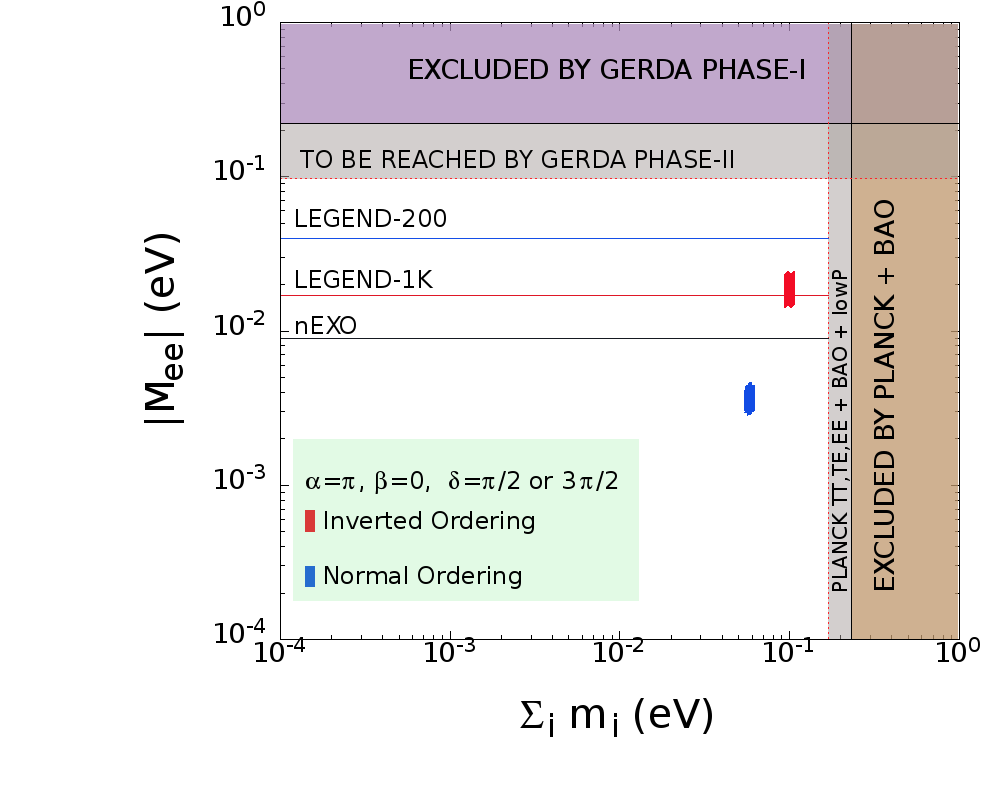}\includegraphics[scale=.22]{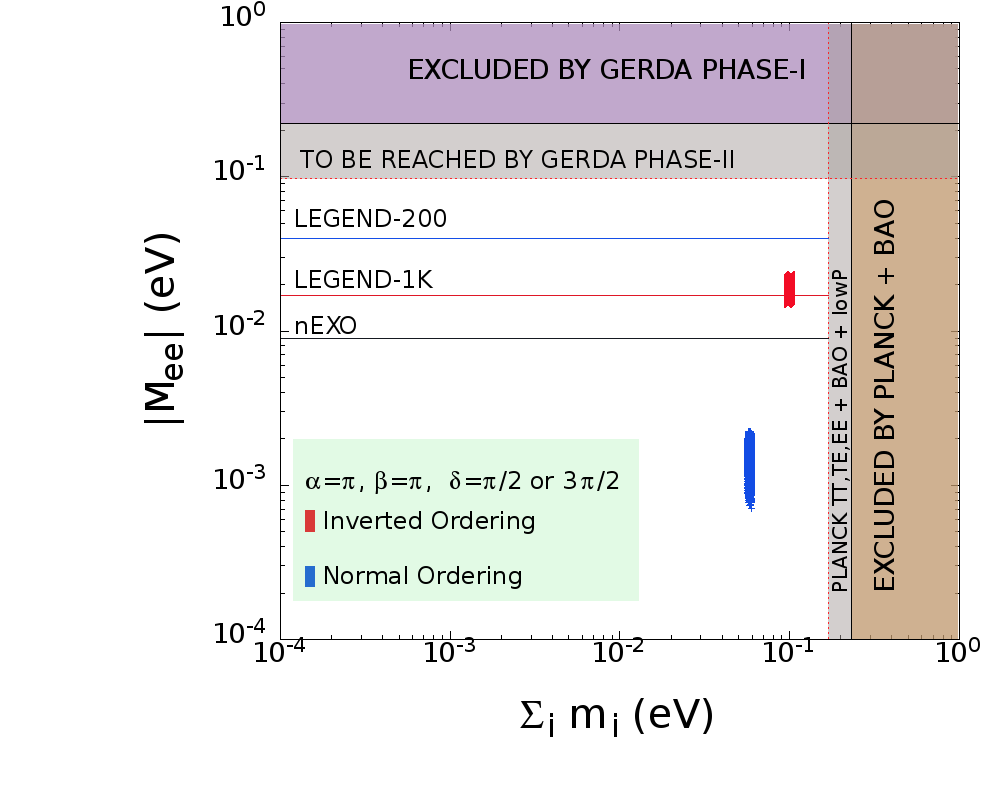}\\

\includegraphics[scale=.22]{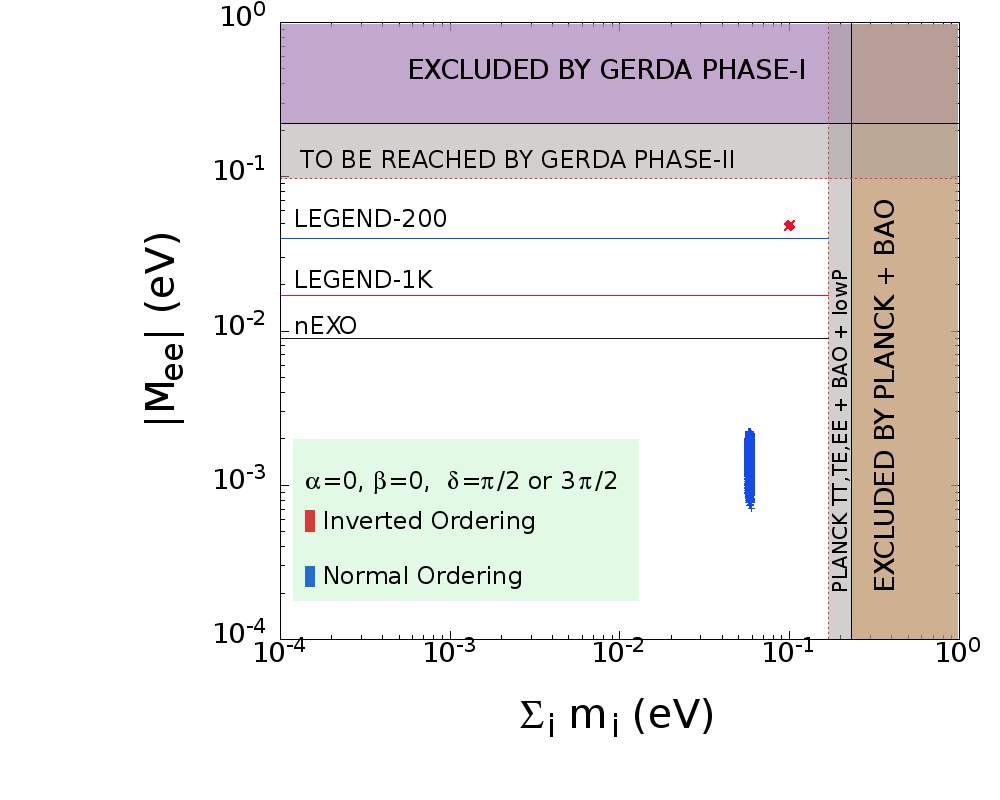}\includegraphics[scale=.22]{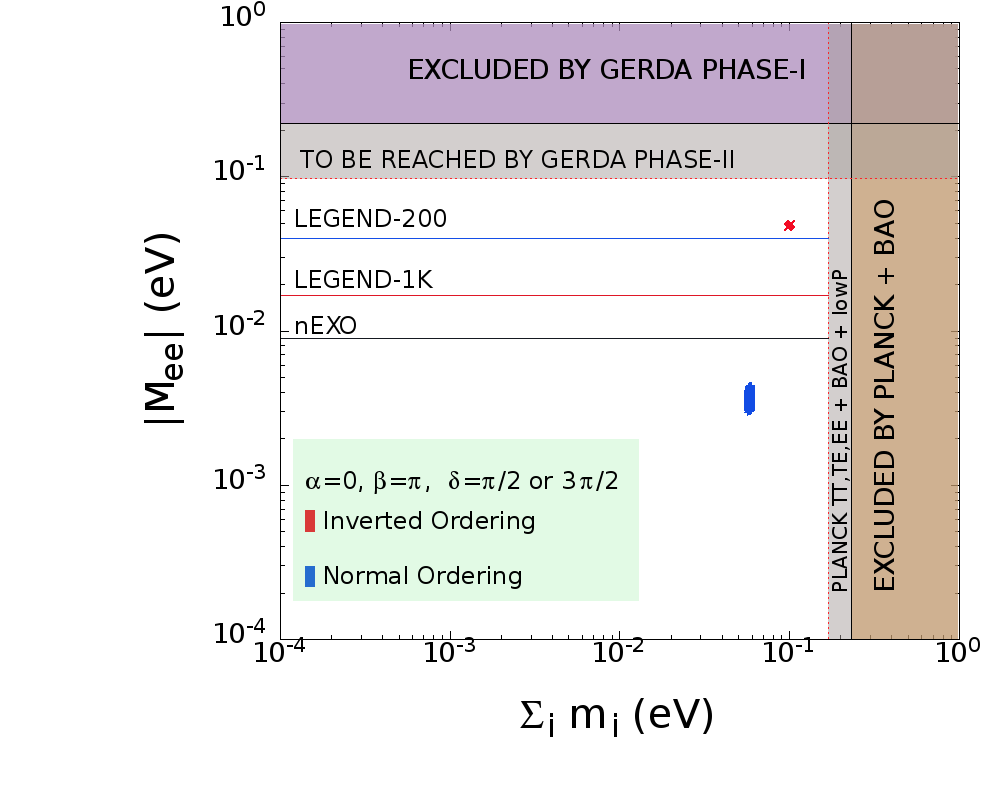}

\caption{Plots: $|M_{ee}|$ vs. $\Sigma_{i}m_i$ for both the mass orderings. }\label{fig2}
\end{figure}

In order to check the viability of our theoretical assumptions and consequent outcomes, we present a numerical analysis in substantial detail. Our method of analysis and organization are as follows. First, we utilize the $3\sigma$ values of the globally fitted neutrino oscillation data presented in  Table \ref{osc1} to  constrain the parameter space in terms of the rescaled parameters defined in (\ref{primed}). For numerical computation, we make use of the exact analytical formulae for the light neutrino masses and mixing angles presented in Ref.\cite{Adhikary:2013bma}. It is seen that, in this complex extended $\mu\tau$ antisymmetry scheme, an appreciable region of the of parameter space could be well fitted within the $3\sigma$ range of the global oscillation data (see Fig.\ref{fig1}) for each of the  mass orderings.
We next discuss the predictions of the present model in the context of the $\beta\beta0\nu$ experiments for both  mass orderings. In order to estimate the value of $Y_B$, we make use of these constrained rescaled parameters with a subtlety. For the computation of  $Y_B$ we need to evaluate the  parameters of $m_D$ (i.e., $a_{1,2}$, $b_{1,2}$) and $M_i$ separately. Since  we have only constrained the rescaled parameters, for a given set of rescaled parameters, there remains a freedom to make various sets of independent choices for the elements of $m_D$ along with $M_i$. Keeping this in mind, we explore two different numerical ways to discuss leptogenesis and its consequent outcomes. First, we choose a specific hierarchical mass spectrum for the RH neutrinos: $M_2/M_1=10^{3}$. Then, for a fixed value of $M_1$, we use the entire parameter space for the rescaled parameters to generate the elements of $m_D$ which are explicitly used to compute the final $Y_B$. This leads to a lower bound on $M_1$ below which $Y_B$ in the observed range cannot be generated. In another approach, instead of taking the entire rescaled parameter space, we focus only on that set of rescaled parameters which corresponds to a positive value of $Y_B$ (the sign of $Y_B$ depends upon the rescaled parameters) and observables that lie near their  best-fit values. Then by varying $M_1$, we generate the corresponding parameters of $m_D$ using (\ref{primed}). Here we consider the same hierarchical scenario for the RH neutrinos as considered in the first approach.  Now, for each value of $M_1$ and the corresponding  parameters of $m_D$, we obtain a value for the final baryon asymmetry $Y_B$. Since $Y_B$ has an observed upper and  a lower bound, we end up with an upper and a lower bound for $M_1$  also. Finally, we provide a numerical discussion regarding the effects  of the heavy neutrino $N_2$ on  the final $Y_B$ as explained analytically in the previous section. We next present the numerical results of our analysis in much more detailed and a systematic  way.\\
\begin{table}[H]
\begin{center}
\caption{Input values fed into the analysis\cite{Capozzi:2017ipn}.} \label{osc1}
 \begin{tabular}{|c|c|c|c|c|c|} 
\hline 
${\rm Parameters}$&$\sin^2\theta_{12}/10^{-1}$&$\sin^2\theta_{23}/10^{-1}$ &$\sin^2\theta_{13}/10^{-2}$ &$ \Delta m_{21}^2/10^{-5}$&$|\Delta m_{31}^2|/10^{-3}$ \\
$ $&${}$ &${}$ &$ {}$&$ \rm (eV^2)$&$  \rm (eV^2) $ \\
\hline
$3\sigma\hspace{1mm}{\rm ranges\hspace{1mm}(NO)\hspace{1mm}}$&$2.50-3.54$&$3.81-6.15$&$1.90-2.40$&$6.93-7.96$&$2.411-2.646$\\
\hline
$3\sigma\hspace{1mm}{\rm ranges\hspace{1mm}(IO)\hspace{1mm}}$&$2.50-3.54$&$3.83-6.36$&$1.90-2.42$&$6.93-7.96$&$2.39-2.624$\\
\hline
${\rm Best\hspace{1mm}{\rm fit\hspace{1mm}}values\hspace{1mm}(NO)}$&$2.97$&$4.25$&$2.15$&$7.37$&$2.52$\\
\hline
${\rm Best\hspace{1mm}{\rm fit\hspace{1mm}}values\hspace{1mm}(IO)}$&$2.97$&$5.89$&$2.16$&$7.37$&$2.50$\\
\hline
\end{tabular} 
\end{center} 
\end{table}
As discussed in Sec.\ref{s3}, there are four sets of CP violating phases for the four independent $\tilde{d}$ matrices. Thus we get four different plots for each mass of the orderings of the light neutrinos. In Fig.\ref{fig2} we present the plots of $|M_{ee}|$ vs. the sum of the light neutrino masses ($\Sigma_{i}m_i$) for each mass ordering. Since the lightest neutrino mass is zero in each case, the other two masses ($m_2$ and $m_3$ for normal ordering and $m_2$ and $m_1$ for inverted ordering) get fixed in a very narrow range by the oscillation constraints on $\Delta m_{21}^2$ and $|\Delta m_{23}^2|$. It is evident from Fig.\ref{fig2} that  $|M_{ee}|$ in each plot leads to an upper limit which is beyond the  reach of the GERDA phase-II. However, predictions of our model could  be probed by the combined GERDA + MAJORANA experiments \cite{Abgrall:2013rze}. The sensitivity reach of  other promising experiments such as LEGEND-200 (40 meV), LEGEND-1K (17 meV) and nEXO (9 meV)\cite{Agostini:2017jim} are also shown in Fig.\ref{fig2}. For each case, the entire parameter space corresponding to an inverted neutrino mass ordering   could be ruled out by the nEXO reach. \\

We now come to the numerical discussion of baryogenesis via flavored leptogenesis. As  mentioned in the beginning of this section, we have performed the numerical computation pertaining to leptogenesis in two different ways. In one way, we have taken a particular value of $M_1$ and compute the final $Y_B$ for the entire rescaled parameter space constrained by the oscillation data. In the second way, we have used those values of the rescaled parameters for which the low energy neutrino observables predicted from our model lie close to their best fit 

\noindent
values dictated by the oscillation data in Table \ref{osc1}. To facilitate this purpose, we define a variable $\chi^2$ that measures the deviation of the parameters from their best fit values:  
\begin{equation}
\chi^2= \sum\limits_{i=1}^5 \Big[ \frac{\mathcal{O}_i(th)-\mathcal{O}_i(bf)}{\Delta \mathcal{O}_i} \Big]^2. \label{chi2}
\end{equation}
In \eqref{chi2} $\mathcal{O}_i$ denotes the $i^{th}$ neutrino oscillation observable from among  $\Delta m^2_{21},\Delta m^2_{32},\theta_{12},\theta_{23}$ and $\theta_{13}$ and the summation runs over all such observables. The parenthetical  $th$ stands for  the numerical value of the observable predicted in our model, whereas $bf$ denotes the best fit value (cf. Table \ref{osc1}). $\Delta \mathcal{O}_i$ in the denominator represents the measured $1\sigma$ range of $\mathcal{O}_i$. Primarily for numerical computation, we choose $M_{2}/M_1=10^3$. However, as indicated in the previous section, we also present a detailed discussion  regarding the  sensitivity of $Y_B$ to the chosen hierarchy of $M_i$. Next, we calculate $\chi^2$  as a function of the primed parameters for their entire constrained range. Then, for a fixed value of $M_1$, we  choose that set of rescaled parameters which corresponds to the minimum value of $\chi^2$ and a positive value of $Y_B$. For that particular $\chi^2$ and the corresponding  set of rescaled parameters, we are then able to generate a large set of elements of $m_D$ by varying $M_1$ over a wide range and can calculate $Y_B$ for each value of $M_1$. An organized discussion is given in what follows.\\

\begin{figure}[H]
\includegraphics[scale=.4]{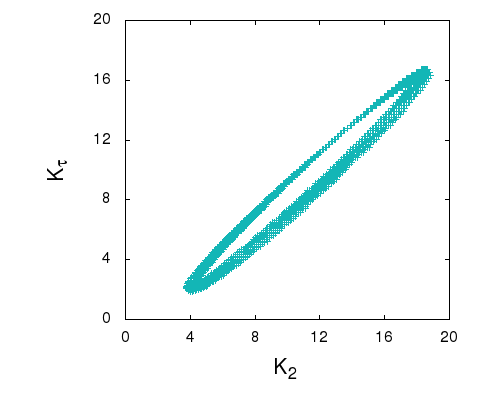}\includegraphics[scale=.4]{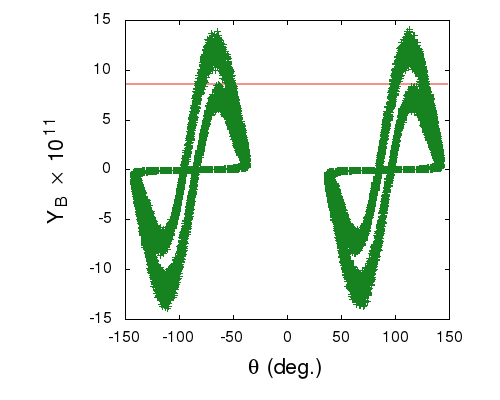}\\
\includegraphics[scale=.4]{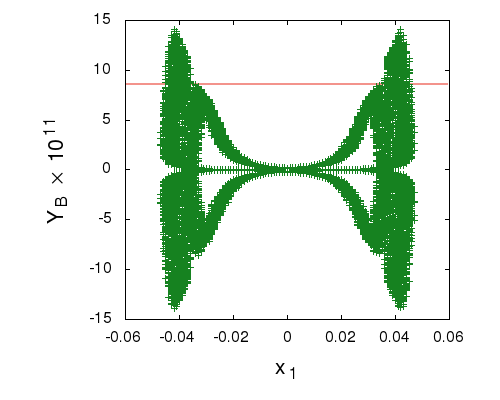}\includegraphics[scale=.4]{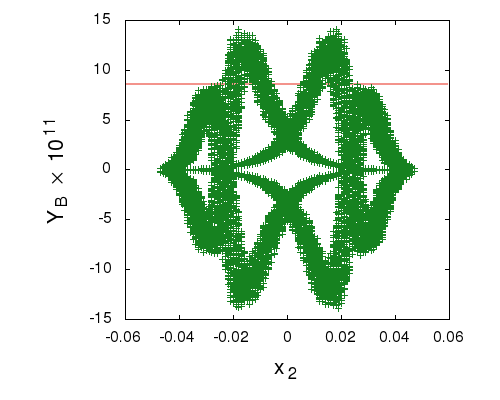}\\
\includegraphics[scale=.4]{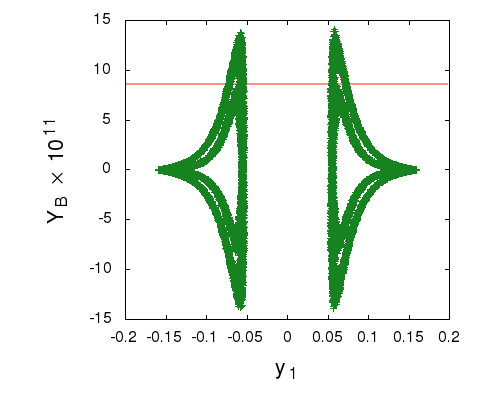}\includegraphics[scale=.4]{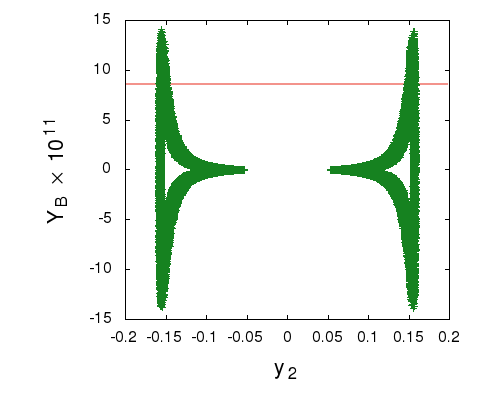}
\caption{The first figure the left panel shows the ranges for the washout parameters. Rest of the plots represent the variation of $Y_B$ with the rescaled parameters for a representative value $M_1=10^{11}$ GeV.}\label{fig3}
\end{figure}

\noindent
\textbf{Computation of $Y_B$ for a normal mass ordering of light neutrinos:}\\
\paragraph{$\bf{M_{1}<{10}^{9}}$ GeV:} In this regime, all three lepton flavors $(e,\mu,\tau)$ are distinguishable. Since $\varepsilon_1^{e}=0$, we need to individually evaluate $\varepsilon_1^{\mu,\tau}$ only. However, due to the imposed $\mu\tau$ antisymmetry, two washout parameters $\tilde{m}_\mu$ and $\tilde{m}_\tau$ would be equal. Thus on account of the relation in (\ref{YBsign}), the final baryon asymmetry $Y_B$ vanishes.
\noindent
\paragraph{$\bf{{10}^{9}\,\,{\rm{\bf{{\rm GeV}}}}<M_{1}<{10}^{12}}$ GeV:} For the evaluation of $Y_B$ here, we have to look first at the washout parameters $K_\tau$ and $K_2=K_e+K_\mu$. As shown in the first plot in the left panel of Fig.\ref{fig3}, the entire allowed range of these parameters  prefers to lie in  $K_\tau$,$K_2>$  1 region. Thus the  efficiency factor in (\ref{tflv})  can be  written in a strong wash-out scenario\cite{Abada:2006ea} as
\bea
\eta(\tilde{m}_\alpha)=\Big[\Big(\frac{0.55\times10^{-3}}{\tilde{m}_\alpha}\Big)^{1.16}\Big],
\eea
where $\alpha=\tau,2$. As elaborated in the previous section,  the assumed strong hierarchy of RH neutrinos makes  the second RHS term in (\ref{m1m2})  much smaller than the first term. Hence  the final CP asymmetry could be simplified to the form as in (\ref{epssim}) so that the final $Y_B$ in (\ref{YBtau}) is practically proportional to the free parameter $M_1$. Now  for a fixed value of $M_1$, we compute $Y_B$ for the entire rescaled parameter space. In Fig.\ref{fig3}, the variation of $Y_B$ with the rescaled parameters is shown for a representative value of $M_1=10^{11}$ GeV. Any further lowering of the value of $M_1$ would cause these plots  (except the first plot in the left panel) to  condense along $Y_B$ axis due to the addressed proportionality of $Y_B$ with $M_1$. Thus, below a certain value of $M_1$, one would end up with a value for $Y_B$ which is below the lower end $8.55\times 10^{-11}$ of the observed range for the latter. We find this lower bound on $M_1$ to be $6.21\times 10^{10}$ GeV for which the peak of a $Y_B$ vs $\theta,x_{1,2},y_{1,2}$ curve in Fig.\ref{fig3} just touches the red stripe that represents the experimental observed range of $Y_B$.\\

Next, we concentrate on the other way which is a search for a set of rescaled parameters that corresponds to the low energy neutrino observables close to their best fit values and hence the minimum value of $\chi^2$. For this purpose, we take a particular set from the rescaled parameter space, calculate the corresponding $\chi^2$ using (\ref{chi2})  and then compute $Y_B$. We have found that  $\chi_{min}^2$ should be $0.397$ for $Y_B$ to be positive. A complete data set of the rescaled parameters and corresponding values of the observables are tabulated in Table \ref{g2nrchi} for $\chi^2_{min}=0.397$.

\begin{table}[H]
\caption{Parameters and observables corresponding  $\chi^2=0.397$ for normal mass ordering.}
\label{g2nrchi}
\begin{center}
 \begin{tabular}{ |c|c|c|c|c|c| } 
 \hline
 $x_1$ & $x_2$ & $y_1$ & $y_2$ & $\theta$ & $\chi_{min}^2$ \\ \hline
 $-0.040$ & $-0.014$ & $-0.01$ & $0.155$ & $114^0$ & $0.397$\\ \hline
\hline
\hline
\multicolumn{2}{|c|}{observables} & $\theta_{13}$ & $\theta_{12}$ &  $\Delta m_{21}^2\times 10^5$ &$|\Delta m_{31}|^2 \times 10^3$\\
\hline
\multicolumn{2}{|c|}{$\chi_{min}^2=0.397$} & $8.42^0$ & $33.04^0$  & $7.47~ {\rm (eV)}^2$ &$2.55~{\rm (eV)}^2$\\
\hline
\end{tabular}
\end{center}
\end{table}

\begin{figure}[H]
\begin{center}
\includegraphics[scale=.45]{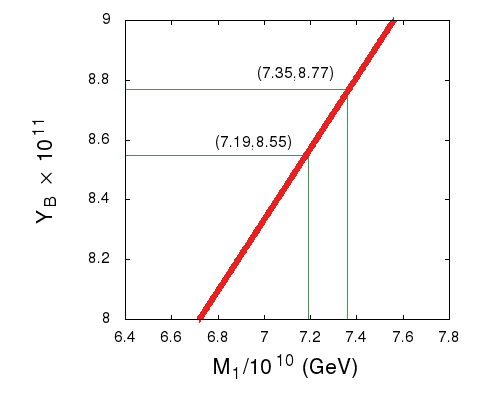}
\caption{$Y_B$ vs. $M_1$ curve corresponding to $\chi^2_{min}=0.397$ for a normal mass ordering of the light neutrinos}\label{fig4}
\end{center}
\end{figure}

Given the rescaled data set for the $\chi^2_{min}$, $M_1$  is varied widely to secure $Y_B$ in the observed range. For each value of $M_1$, a set of values of the  parameters in the elements  of $m_D$ is generated. The final $Y_B$ is then calculated for each value of $M_1$ and the corresponding parameters of $m_D$. A careful surveillance of the plot in  Fig.\ref{fig4} leads to the conclusion that we can obtain an upper and a lower bound on $M_1$ corresponding to the observed constraint on $Y_B$. In order to realize  this  fact  more clearly,   two straight lines have been drawn parallel to the abscissa in the mentioned plot: one at $Y_B=8.55\times10^{-11}$ and the other at $Y_B=8.77\times10^{-11}$.  The values of $M_1$, where the straight lines connect the  $Y_B$ vs $M_1$ curve, yield the allowed upper and lower bounds on $M_1$, namely $(M_1)_{upper}=7.35\times10^{10}$ GeV and  $(M_1)_{lower}=7.19\times10^{10}$ GeV.  Again, the near linearity of the $Y_B$ vs. $M_1$ curve in  Fig.\ref{fig4} follows from the previously explained approximate proportionality of $Y_B$ with $M_1$. One might also ask about the narrow range for $M_1$ as observed in  Fig.\ref{fig4}. Note that in this plot we have presented our result for a particular set of  rescaled parameters (with $\chi^2_{min}=0.397$). In principle, one could take the entire rescaled parameter space of our model and compute the corresponding results on $Y_B$ and $M_1$ for each set of the mentioned parameters. In that case  the range of $M_1$ would not be as narrow as shown in Fig.\ref{fig4}.
\paragraph{$\bf{M_{1}>{10}^{12}}$ GeV:} In this regime $Y_B$ is zero since the flavored sum CP asymmetry parameter $\sum_{\alpha}\varepsilon_1^\alpha$ vanishes. Obviously,  $Y_B$ might be generated in this regime also if one consider small breaking of CP symmetry in the neutrino sector as discussed in Ref.\cite{Hagedorn:2016lva}.\\

\noindent
\textbf{Computation of $Y_B$ for an inverted mass ordering of light neutrinos:}\\

\noindent
In this case also the observed range of $Y_B$ cannot be generated for $M_{1}<{10}^{9}$ GeV and $M_{1}>{10}^{12}$ GeV owing to  reasons similar to those explained in the case of a normal ordering. However, we find that in the case of an inverted ordering, $Y_B$ cannot be generated in the observed range even if we consider a $\tau-$flavored regime, i.e., $10^9$ GeV $<M_1<10^{12}$ GeV. Numerically, for a value $M_1 =9.9\times10^{11}$ GeV, $Y_B$  is computed to be $Y_B=8.20\times 10^{-11}$. Thus from a hierarchical leptogenesis perspective, an inverted mass ordering is disfavored in our model with a complex (CP) extended antisymmetry.\\

\noindent
\underline{The effect of $N_{2}$ on $Y_B$}\\

\begin{figure}[H]
\includegraphics[scale=.4]{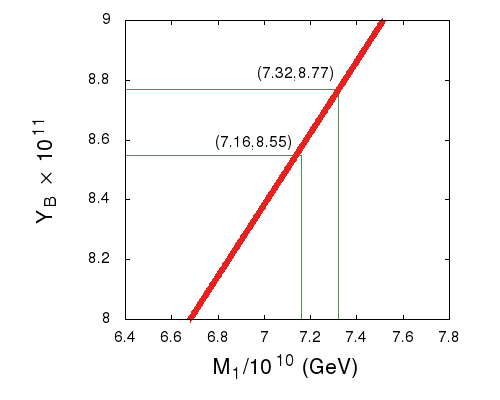}\includegraphics[scale=.4]{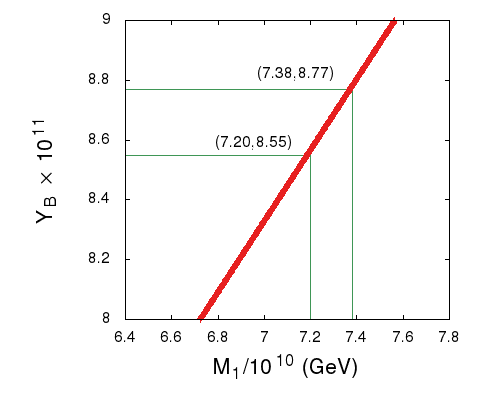}
\caption{$Y_B$ vs. $M_1$ plots corresponding to $\chi^2_{min}=0.397$ for the normal mass ordering of the light neutrinos. The plot in the left is for $M_2/M_1=10^{2}$ and the plot in the right is for $M_2/M_1=10^{4}$.}\label{ybm1h}
\end{figure}

\begin{figure}[H]
\begin{center}
\includegraphics[scale=.45]{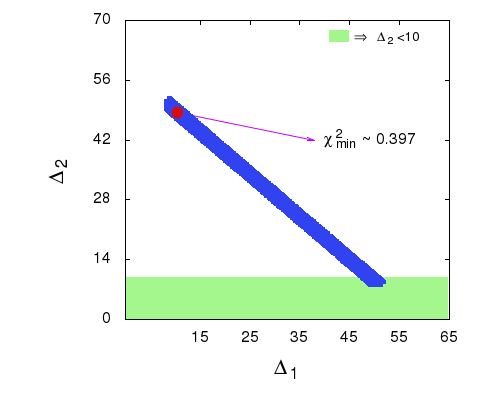}
\caption{washout parameters for $N_2$ leptogenesis}\label{fig5}
\end{center}
\end{figure}

\noindent
As mentioned in the previous section, there are two different ways in which the heavy RH neutrino $N_2$ might affect the final value of $Y_B$. In the first, which we name as the indirect effect,  the final $Y_B$ becomes practically insensitive to the mass of $N_2$ since the second term is suppressed compared to the first term in (\ref{m1m2}). Now $\varepsilon_1^\mu$ can be written in a simpler form which is independent of $M_2$. c.f, (\ref{epssim}); hence it does not depend the mass ratio $M_2/M_1$. However, for a precise computation of $Y_B$, we need to consider the term  neglected in (\ref{m1m2}); that in turn motivates us to perform a quick check of the RH neutrino mass hierarchy sensitivity of the produced value of $Y_B$. For this purpose, in addition to the standard hierarchical case, i.e. $M_2/M_1=10^{3}$,   we calculate $Y_B$ for two other different mass hierarchical schemes, $M_{2}/M_1=10^2$ and $M_{2}/M_2=10^4$.   From  Fig.\ref{ybm1h} we can infer that  though the chosen mass ratios of the RH neutrinos are altered, changes in the lower and upper bounds on $M_1$ are practically insignificant.  For the allowed normal light neutrino mass  ordering, the variation of $Y_B$ with $M_1$ for different mass ratios of the RH neutrinos has been presented  in Table \ref{t8}.
\begin{table}[H]
\begin{center}
\caption{Lower and upper bounds on $M_1$ for different mass ratios of the RH neutrinos.} \label{t8}
 \begin{tabular}{|c|c|c|c|} 
\hline 
 \multicolumn{4}{|c|}{\cellcolor{gray!30}Case-I: Normal light neutrino ordering}\\
${\rm Hierarchies~\rightarrow }$&$M_{2}/M_1=10^2$&$M_{2}/M_1=10^3$ &$M_{2}/M_1=10^4$\\
\hline
${\rm Upper~bound~(GeV)}$&$7.32\times 10^{10}$&$7.35 \times 10^{10}$ &$7.38 \times 10^{10}$ \\
\hline
${\rm Lower~bound~~(GeV)}$&$7.16 \times 10^{10}$&$7.19 \times 10^{10}$&$7.20 \times 10^{10}$\\
\hline
\end{tabular} 
\end{center} 
\end{table}
It is obvious from the entries of Table \ref{t8} that a slight difference in the upper and lower bounds on $M_1$ in a particular column, as compared to the other column, arises due the  dependence of $M_2$ on the second term in (\ref{m1m2}). For a fixed value of $M_1$, the contribution from the second term in (\ref{m1m2}) is larger for $M_{2}/M_1=10^{2}$ and smaller for $M_{2}/M_1=10^{4}$, as compared to the standard $M_{2}/M_1=10^{3}$ case. Hence for  $M_{2}/M_1=10^{2}$, the slope of the $Y_B$ vs. $M_1$ curve is larger  than for $M_{2}/M_1=10^{3}$. Consequently, for the allowed range of $Y_B$, both  the upper and the lower bounds get slightly left shifted on the $M_1$-axis (compared to the standard $M_{2}/M_1=10^{3}$ case).  Proceeding in the same way, we obtain  somewhat right shifted bounds for $M_{2}/M_1=10^{4}$ case.\\
In contrast, in the direct effect,  any asymmetry produced by $N_2$ survives provided the conditions   $\Delta_1\gg 1\hspace{1mm} {\rm and}\hspace{1mm}\Delta_2 \not\gg 1$, cf. (\ref{n2cond}), are satisfied.  From Fig.\ref{fig5} we observe that the 

\noindent
allowed  parametric region  prefers large values of $\Delta_2$ in excess of 10 except at the bottom (green band). Thus the condition $\Delta_2 \not\gg 1$ is violated in most of the region. Moreover the $\chi^2_{min}=0.397$, for which we calculate final $Y_B$ strongly violates $\Delta_2 \not\gg 1$ condition. A tiny amount of parameter space with $\Delta_2<10$ corresponds to values of $\chi^2$ above 0.9 which is much higher than $\chi^2_{min}$ for which we compute $Y_B$ in the observed range. Therefore, in our final result, any direct effect of $N_2$ is not significant. Note that there is nothing special about $\chi^2=0.9$. The issue we are trying to address here, is that there are indeed some data points in the model parameter space for which the conditions for $N_2$ leptogenesis could be satisfied. However, the minimum value of $\chi^2$ for those data sets is 0.9. This means that the corresponding observables are  away from their best-fit values (though well within 1$\sigma$) and thus the obtained  bounds on $M_1$ (e.g. Fig.\ref{fig4}) will not be affected by $N_2$ leptogenesis. However, if one goes beyond $\chi^2\approx 0.9$, the asymmetry produced by $N_2$ could play a crucial role.  \\

We would like to conclude this section by comparing our results 
on leptogenesis with those obtained earlier in previous  
literature in case of a  $\mu\tau$ flavored CP symmetry. Existing 
references such as \cite{Mohapatra:2015gwa,Chen:2016ptr,Hagedorn:2016lva} also discuss leptogenesis within the framework of residual CP symmetry (in particular $\rm CP^{\mu\tau}$) and point out the nonoccurrence of unflavored leptogenesis and only the viability of the $\tau-$flavored scenario similar to our proposal of a exact $\mu\tau$ antisymmetry  in the neutrino sector. However, the final numerical analysis is different  from our case. In particular, all the mentioned references mainly focus on the three neutrino case where one cannot fix the Yukawa couplings only with the oscillation data. Thus any final result on 
leptogenesis requires other assumptions to constrain all the Yukawas. We focus on the  two RH neutrino case, namely the minimal seesaw mechanism, where the entire Yukawa parameter (rescaled by RH neutrino masses) space  could be 
constrained by the neutrino oscillation data. Hence all the results obtained, 
in particular for the RH neutrino masses, are exactly dictated by the oscillation data. For a hierarchical RH mass spectrum, Ref.\cite{Chen:2016ptr} shows a variation of $Y_B$ with a single model parameter for a fixed value of $M_1$ and best-fit values of the oscillation parameters. However, here we focus on the bounds on $M_1$ for the entire parameter space as well as the for the parameter set that corresponds to the observables which lie near to their best-fit values. For the first case, we obtain a  lower bound on $M_1$ while in the other, we obtain an upper as well as  a lower bound on $M_1$. In addition, we have done a thorough study of the RH neutrino hierarchy sensitivity of the final $Y_B$ and showed the possible changes in the  bounds on the lightest RH neutrino $M_1$ for three different RH neutrino hierarchical mass spectra. We have also showed that, for this minimal seesaw  with a complex $\mu\tau$ antisymmetry, the inverted mass ordering is not a viable option as far as hierarchical leptogenesis is concerned.  
We are not within the  framework of a Grand Unified Theory (GUT) such as SO(10), where the lepton asymmetry generated by the next to light RH neutrino ($N_2$), is a natural requirement to produce correct value of $Y_B$\cite{Akhmedov:2003dg,Di}. Nevertheless, we opt for fast $N_1$ interactions which are responsible for the survival of the lepton asymmetry generated by $N_2$\cite{Engelhard:2006yg}. For this CP symmetric framework we have showed for the first time  that there could be a parameter space left for which $N_2$ leptogenesis might affect the final value of $Y_B$ (though a rigorous study of the $N_2$ leptogenesis is beyond the scope of this paper). Ref.\cite{Mohapatra:2015gwa} concluded that  for the mass regime $M_1<10^9{\rm GeV}$, a resonant leptogenesis is only possible if one considers breaking in $CP^{\mu\tau}$, since in this regime the muon and tauon washout parameters are of equal strength. In our proposal of $CP^{\mu\tau A}$ also, this conclusion is true. However, as we show in the appendix, unlike  the hierarchical RH neutrino mass spectrum, RH neutrinos with a mild hierarchy could also result in a successful leptogenesis for $M_1\approx10^9{\rm GeV}$. We showed that an inverted mass ordering could then be a viable option for a successful leptogenesis. We also comment on the strength of the mild hierarchy by solving numerically the formulae for $Y_B$ within the framework of  flavor diagonal RH neutrinos. In our analysis, the sign of $Y_B$ depends upon the Yukawa parameters. In 
this context we refer to\cite{Hagedorn:2016lva} which shows how, within the 
framework of a CP symmetry, the sign of $Y_B$ depends upon the observables. 

\section{Concluding comments and discussion}\label{s8}
In this paper the complex (CP) extension of $\mu\tau$ antisymmetry has been 
shown to yield a $M_\nu^{{CP}^{\mu\tau A}}$ which is simply related to 
$M_\nu^{{CP}^{\mu\tau }}$ - the Majorana mass matrix from the complex (CP) 
extension of $\mu\tau$ symmetry with both having identical phenomenological 
consequences. These phenomenological consequences of $CP^{\mu\tau A}$ have been 
worked out within a minimal seesaw scheme with two strongly or mildly 
hierarchical RH neutrinos $N_1$ and $N_2$. We have further investigated 
baryogenesis via leptogenesis in this scenario and derived upper and lower 
bounds on the mass of $N_1$.      
\vskip 0.1in
\noindent
To summarize, we have proposed a new idea, namely a complex extended 
$\mu\tau$ antisymmetry, pertaining to the neutrino sector and have worked out its consequences. Unlike the real $\mu\tau$ antisymmetry, we  envisage there is no need for any breaking of it in the neutrino sector. Atmospheric neutrino mixing  is predicted to be maximal ($\theta_{23}=\pi/4$) in this scheme while the solar and reactor mixing angles  ($\theta_{12}$ and $\theta_{13}$ respectively) can be fit to their observed values. Neutrino masses get generated via the minimal seesaw mechanism with two heavy right-chiral neutrinos. The lightest neutrino is predicted to be massless while the two other neutrino masses can be fit to the observed range of values of $|\Delta m_{32}^2|$ and $\Delta m_{12}^2$ both for a normal and an inverted mass ordering. Concrete predictions are made for neutrinoless double beta decay: the ongoing experiments are not expected to observe it  though the planned nEXO experiment may have a chance to do so. Finally, we have made a detailed quantitative examination of baryogenesis via leptogenesis in our scheme including the indirect and direct effects of the heavier RH neutrino $N_2$. $\tau$-flavored leptogenesis with a normal mass ordering turns out to be the only viable possibility that can generate $Y_B$ in the observed range in a hierarchical leptogenesis scenario.
\section*{Acknowledgement}
The work of RS and AG is supported by the Department of Atomic Energy (DAE), Government of India. The
work of PR has been supported by the Indian National Science Academy.
\vskip 0.1in
\appendix
\section{Discussion of the case with mildly hierarchical RH 
neutrinos}

In the text  we have dealt with a strongly hierarchical RH neutrino mass 
spectrum and found  only  
the $\tau-$flavored regime to be viable in producing the correct $Y_B$ for 
a normal light neutrino mass ordering. Since in our chosen basis \cite{Chen:2016ptr}, 
RH neutrinos are nondegenerate, it  would also be interesting to study 
leptogenesis with a mildly hierarchical including a  quasidegenerate  
$N_R$ mass spectrum. We will see later in this discussion that RH neutrinos 
which are not strongly hierarchical might obliterate all the 
new bounds on $M_1$ that we obtained earlier.
\vskip 0.1in
\noindent
In general, a quasidegenerate RH neutrino mass spectrum is considered 
for studying leptogenesis in a low energy seesaw scenario 
(resonant leptogenesis\cite{Pilaftsis:2003gt}); here the RH neutrinos could have 
masses $\mathcal{O}({\rm TeV})$. 
However, in our analysis, we cannot lower the RH neutrino masses 
below $10^{9}$ GeV, since that would correspond to the fully flavored 
regime where the two washout parameters $\tilde{m}_{\mu}$ and 
$\tilde{m}_{\tau}$  are the same due to the imposed $\mu\tau$ antisymmetry, 
thereby implying  a vanishing $Y_B$ cf.(\ref{YBsign}). However, depending on 
the chosen mild mass splitting of the RH neutrinos, we can lower the 
lightest RH neutrino mass down to $10^{9}$ GeV below which the muon 
charged lepton flavor equilibriates. In scenarios where  the RH neutrinos 
are not strongly hierarchical, instead of (\ref{asymp}), 
it is useful to use the  general formula for the CP asymmetry parameter\cite{Pilaftsis:2003gt} 
$\varepsilon_i^\alpha$ as 
\bea
\varepsilon^\alpha_i
&=&\frac{1}{4\pi v^2 h_{ii}}\sum_{j\ne i} {\rm Im}\{h_{ij}
({m_D})_{i\alpha} (m_D^*)_{j\alpha }\}
\left[f(x_{ij})+\frac{\sqrt{x_{ij}}(1-x_{ij})}
{(1-x_{ij})^2+{{h}_{jj}^2}{(16 \pi^2 v^4)}^{-1}}\right]\nonumber\\
&+&\frac{1}{4\pi v^2 {h}_{ii}}\sum_{j\ne i}\frac{(1-x_{ij})
{\rm Im}\{{h}_{ji}({m_D})_{i\alpha} (m_D^*)_{j\alpha}\}}
{(1-x_{ij})^2+{{h}_{jj}^2}{(16 \pi^2 v^4)}^{-1}}.\label{ncp}
\eea 
Note that, unlike (\ref{asymp}) the above equation is valid for degenerate RH 
neutrinos also.
\vskip 0.1in
\noindent
Taking into account the contribution from both the RH neutrinos, 
we have performed a numerical study to find the final $Y_B$ for the 
lowest allowed value of $M_1 (=10^{9}{\rm GeV})$. 
It turns out that for a normal light neutrino mass ordering, 
$M_2$ could at most be $\approx 17.5M_1$ to produce the 
observed lower bound $8.55\times 10^{-11}$ of $Y_B$ cf.(\ref{Ybob}). One can see 
that the obtained mass spectrum is {\bf fairly hierarchical} though the 
hierarchy is not very strong.  
Of course any number smaller than 17.5 would result in  an enhancement of 
the produced CP asymmetry. Thus the observed range of $Y_B$ could be 
generated with a quasidegenerate RH mass spectrum too. 
Interestingly, an inverted light neutrino mass ordering which is 
disfavoured for a 
strongly hierarchical RH neutrino mass spectrum is now a perfectly viable 
scenario since we relax the strong hierarchy assumption. 
Again, as in the previous case, i.e., for $M_1=10^{9} {\rm GeV}$, it is 
numerically found that one needs $M_2\leqslant1.8 M_1$ in order to produce 
the observed lower bound on $Y_B$. Note that, unlike in the case of a 
normal light neutrino mass ordering, the RH neutrino mass spectrum here 
favours a {\bf mild hierarchical scenario} as we lower the value of $M_1$. We could also point out that here we have considered the flavor diagonal RH neutrinos  to calculate the asymmetry. Nevertheless, for a resonant leptogenesis scenario, a full flavor-covariant treatment might play an important role\cite{Abada:2006ea2}.
\vskip 0.1in
\noindent
As a concluding remark, we may mention once again that, owing to the 
imposed symmetry, a fully flavored leptogenesis is not possible 
for $M_1<10^{9}{\rm GeV}$  even if we consider  strongly degenerate 
RH neutrinos. Nevertheless, a small breaking of the symmetry\cite{Hagedorn:2016lva}, 
or somewhat a more moderate version of the symmetry 
such as the scaling ansatz \cite{CP1}
will cause a deviation from  
$\tilde{m}_{\mu}$=$\tilde{m}_{\tau}$ cf. (\ref{YBsign}) and will imply a 
nonvanishing $Y_B$. In such cases leptogenesis with heavily 
degenerate RH neutrinos (resonant leptogenesis) could be an 
interesting topic to study.

\end{document}